\documentclass[twocolumn,aps,prb,superscriptaddress]{revtex4-1}

\usepackage{amsmath}
\usepackage{graphicx}
\usepackage{color}
\usepackage{url}
\usepackage[left=0.75in, right=0.75in,top=0.75in,bottom=1.5in]{geometry}

\begin{document}

\title{Spin-density fluctuations and the fluctuation-dissipation theorem in 3\emph{d} ferromagnetic metals}

\author{A. L. Wysocki}
\email{alexwysocki2@gmail.com}
\affiliation{Ames Laboratory, Ames, IA 50011, USA}

\author{V. N. Valmispild}
\affiliation{Institute of Theoretical Physics, University of Hamburg, 20355 Hamburg, Germany}
\affiliation{The Hamburg Centre for Ultrafast Imaging, Luruper Chaussee 149, 22761 Hamburg, Germany}

\author{A. Kutepov}
\altaffiliation[Present address: ]{Brookhaven National Laboratory, Upton, NY 11973-5000, USA}
\affiliation{Ames Laboratory, Ames, IA 50011, USA}

\author{S. Sharma}
\affiliation{Max Planck Institute of Microstructure Physics, Weinberg 2, D-06120 Halle, Germany}

\author{J. K Dewhurst}
\affiliation{Max Planck Institute of Microstructure Physics, Weinberg 2, D-06120 Halle, Germany}

\author{E. K. U. Gross}
\affiliation{Max Planck Institute of Microstructure Physics, Weinberg 2, D-06120 Halle, Germany}

\author{A. I. Lichtenstein}
\affiliation{Institute of Theoretical Physics, University of Hamburg, 20355 Hamburg, Germany}
\affiliation{The Hamburg Centre for Ultrafast Imaging, Luruper Chaussee 149, 22761 Hamburg, Germany}

\author{V. P. Antropov}
\affiliation{Ames Laboratory, Ames, IA 50011, USA}

\date{\today}

\begin{abstract}
Spatial and time scales of spin density fluctuations (SDF) were analyzed in 3\emph{d} ferromagnets using \emph{ab initio} linear response calculations of complete wavevector and energy dependence of the dynamic spin susceptibility tensor. We demonstrate that SDF are spread continuously over the entire Brillouin zone and while majority of them reside within the 3\emph{d} bandwidth, a significant amount comes from much higher energies. A validity of the adiabatic approximation in spin dynamics is discussed. The SDF spectrum is shown to have two main constituents: a minor low-energy spin wave contribution and a much larger high-energy component from more localized excitations. Using the fluctuation-dissipation theorem (FDT), the on-site spin correlator (SC) and the related effective fluctuating moment were properly evaluated and their universal dependence on the 3\emph{d} band population is further discussed.
\end{abstract}

\maketitle

\section{Introduction}

The physics of spin density fluctuations\cite{MoriyaBook,TakahashiBook} (SDF) in metallic magnets is very rich and complex. SDF determine the magnetic excitation spectrum and play an important role in the magnetic dynamics. In addition, they can strongly affect numerous static magnetic and nonmagnetic properties at zero and finite temperatures.\cite{SolontsovBook,KimBook} SDF can be especially important near phase transitions points where they can stabilize new ground states.\cite{QCPReview}

The key quantity characterizing SDF in metals is the spin correlator (SC) which represents the equal-time on-site connected spin correlation function and plays a crucial role in spin fluctuation theories (see, e.g., Ref. \onlinecite{TakahashiBook}). According to the fluctuation-dissipation theorem\cite{Kubo} (FDT), SC can be evaluated by integrating the imaginary part of the dynamic spin susceptibility over all wavevectors and energies. Such integration is, however, a highly nontrivial task both for the experiment and theory which makes reliable calculations of SC in real materials very difficult.

For instance, experimentally, SDF are traditionally studied using the neutron scattering technique. This method may be used to obtain the imaginary part of the dynamic spin susceptibility for certain points in Brillouin zone (BZ) when energies are below $\sim$0.3-0.4 eV.\cite{Dai} While from this information SC has been evaluated for many systems,\cite{SolontsovBook} such estimates are not very accurate due to small number of wavevectors and limited energy range used in the calculations. In addition, despite the clear itinerant nature of magnetic metals, in most studies the experimental results have been compared with the conclusions of a localized spin model (Heisenberg). Independently, fast and ultrafast spin dynamical experiments also detect the presence of SDF at very different frequency ranges.\cite{exp,ultrafast} However, those studies are usually not related, and so far, no consistent experimental measurements of the full SDF spectra in a wide energy range have been performed.  

Theoretically, SDF in real materials can be explored from first principles using linear response technique based on density functional theory\cite{Savrasov,Staunton,Costa,Buczek,Bergara,Lounis} or many-body perturbation methods.\cite{Blugel,Schilfgaarde} However, these calculations considered only limited energy and wavevector ranges. Consequently, proper evaluation of SC for magnetic metals is currently missing in the literature.

In addition to linear response studies of SDF, numerous theories including SDF in calculations of ground state or thermodynamic properties of materials were developed. These methods, however, have also been restricted to narrow bands energy scale and/or limited parts of the BZ. In particular, spin fluctuation models which were widely used to study effects of SDF in 3\emph{d} ferromagnets\cite{Shimizu,Lonzarich,Solontsov,SolontsovBook,Solontsov2,Antropov} employ long wavelength and low frequency approximations. On the other hand, dynamical mean-field theory\cite{Kotliar} (DMFT) or single-site many-body perturbation theory\cite{Zein} include only pure intra-atomic SDF on a limited energy range. These approximations can especially affect accuracy of SC values calculated using DMFT.\cite{Lichtenstein,Yin,Igoshev} While the above mentioned approaches have been successful in the description of many systems, their essentially adjustable nature and uncontrollable approximations do not allow us to understand the relative roles of the different spatial or time scales of SDF in determining materials properties.

Clearly, a comprehensive study of the full structure of SDF in metallic magnets is necessary for a rigorous evaluation of SC. In addition, such analysis would provide an important insight about the scales of SDF that should be included in electronic structure calculations. We recently addressed this issue in 3\emph{d} paramagnetic metals\cite{Wysocki} where it was shown that itinerant SDF are present throughout BZ and a wide energy range. Using FDT SC was evaluated resulting in a strong effective fluctuating moment that was found to be determined solely by the 3\emph{d} band population. For ferromagnetic metals, however, it is unclear how local moments interact with such itinerant SDF. Theories based on the localized Heisenberg model, which are very successful in magnetic insulators, are no longer applicable because of this intrinsic itinerancy. Therefore, a proper quantum-mechanical treatment is crucial in order to establish quantitative description of SDF in magnetic metals.

A primary goal of this paper is to present such analysis by using realistic electronic structure calculations. We focus on prototype 3\emph{d} ferromagnets including Fe, Co (fcc), and Ni, where the degree of moment localization is changing gradually. Using two independent computational techniques, we determine the strength and the character of such SDF as well as establish their spatial and energy scales. SC is properly evaluated using FDT and the dependence of the results on the 3\emph{d} band populations is studied.

\section{Method}

\subsection{SDF formalism}

SDF in solids are described by the imaginary part of the dynamic spin susceptibility tensor 

\begin{eqnarray}
\nonumber
\chi^{\alpha\beta}(\mathbf{r},\mathbf{r}',\mathbf{q},\omega)= -\frac{i}{\hbar}\sum_{\mathbf{R}}e^{-i\mathbf{q}\cdot\mathbf{R}} \\
\times\int_0^\infty dt\langle[\hat{s}_{\alpha}(\mathbf{r}+\mathbf{R},t),\hat{s}_{\beta}(\mathbf{r}')]\rangle e^{i(\omega+i\eta)t}.
\label{chi}
\end{eqnarray}

Here, $\alpha,\beta=x,y,z,0$ denote components of the tensor, $\mathbf{r}$ and $\mathbf{r}'$ are position vectors inside the crystal unit cell, $\mathbf{q}$ is the wavevector in the Brillouin zone, $\omega$ is the frequency, $\mathbf{R}$ is the lattice vector, $\langle...\rangle$ denotes the thermal and quantum-mechanical expectation value, $\eta\rightarrow 0^+$, and $\hat{s}_\alpha(\mathbf{r},t)$ is the density operator when $\alpha=0$, otherwise it is the $\alpha$ component of the spin density operator. For collinear magnetic states and in the absence of spin-orbit coupling, many of the tensor elements are zero. In particular, if one chooses the $z$ axis along the magnetization direction (or sublattice magnetization in the case of antiferromagnets), the susceptibility tensor in the matrix notation becomes

\begin{equation}
    \check{\chi}=\left(\begin{array}{cccc}
     \chi^{xx} & \chi^{xy} & 0         & 0         \\
    -\chi^{xy} & \chi^{xx} & 0         & 0         \\
      0        & 0         & \chi^{zz} & \chi^{0z} \\
      0        & 0         & \chi^{0z} & \chi^{00} \\
    \end{array}\right),
\label{mchi}
\end{equation}

where the dependence on  $\mathbf{r}$, $\mathbf{r}'$, $\mathbf{q}$, and $\omega$ variables is not shown explicitly. It is convenient to express the transverse components ($\chi^{xx}$ and $\chi^{xy}$) in terms of the circular susceptibilities

\begin{eqnarray}
\chi^{+-}=2\left(\chi^{xx}-i\chi^{xy}\right) \\
\chi^{-+}=2\left(\chi^{xx}+i\chi^{xy}\right).
\label{chicirc}
\end{eqnarray}

Note that the transverse components are decoupled from the longitudinal susceptibility ($\chi^{zz}$). On the other hand, $\chi^{zz}$ is coupled to the density response ($\chi^{00}$) through the  spin-density susceptibility function $\chi^{0z}$. 

For SDF analysis, it is often not necessary to resolve intra-atomic fluctuations. Therefore, for each nonequivalent atom it is convenient to introduce the SDF spectral function

\begin{equation}
A^{\alpha\beta}(\mathbf{q},\omega)= -\frac{\hbar}{\pi}\int d\mathbf{r}\int d\mathbf{r}^\prime\ \text{Im}\chi^{\alpha\beta}(\mathbf{r},\mathbf{r} ^{\prime},\mathbf{q},\omega),
\label{ASDF}
\end{equation} 

where $\mathbf{r}$ and $\mathbf{r}'$ variables are integrated over the atomic sphere. Correspondingly, the density of on-site SDF can be defined by integrating $A^{\alpha\beta}(\mathbf{q},\omega)$ over the BZ

\begin{equation}
N^{\alpha\beta}(\omega)=\frac{1}{\Omega_{\text{BZ}}}\int_{\Omega_{\text{BZ}}}d\mathbf{q} A^{\alpha\beta}(\mathbf{q},\omega). 
\label{NSDF}
\end{equation} 

where $\Omega_{\text{BZ}}$ is the BZ volume. In order to better analyze the distribution of SDF in the BZ, one can also consider the partial-$\mathbf{q}$ density of on-site SDF defined as

\begin{equation}
N_{\Omega_{\mathbf{q}}}^{\alpha\beta}(\omega)=\frac{1}{\Omega_{\text{BZ}}}\int_{\Omega_{\mathbf{q}}}d\mathbf{q} A^{\alpha\beta}(\mathbf{q},\omega). 
\label{NrSDF}
\end{equation} 

Here, the integration is over a $\Gamma$-point-centered sphere with the volume $\Omega_{\mathbf{q}}$ ($\Omega_{\mathbf{q}}<\Omega_{\text{BZ}}$). Further, we introduce the on-site number of transverse SDF $n^t(\omega)$ as well as longitudinal SDF $n^l(\omega)$,

\begin{eqnarray}
n^t(\omega) =\frac{1}{2}\int_{0}^{\omega }d\omega'\left[N^{+-}(\omega')+N^{-+}(\omega')\right] \\
\label{ntSDF}
n^l(\omega) =\int_{0}^{\omega }d\omega' N^{zz}(\omega').
\label{nlSDF}
\end{eqnarray}

FDT plays a crucial role in the physics of SDF since it allows to find a number of useful properties that characterize the SDF spectrum. In particular, it can be used to evaluate SC which is defined as an energy integral of the dynamic on-site connected spin correlation function and is an important measure of the strength of SDF in solids. According to FDT, the transverse and longitudinal contributions to the SC are given by

\begin{eqnarray}
\left\langle\mathbf{s}^{2}\right\rangle^t_\omega =\frac{1}{2}\int _{0}^{\omega }d\omega'\coth{(\beta\omega'/2)} \\
\times\left[N^{+-}(\omega')+N^{-+}(\omega')\right],
\label{SCt}
\end{eqnarray}

and

\begin{eqnarray}
\left\langle\mathbf{s}^{2}\right\rangle^l_\omega =\int _{0}^{\omega }d\omega'\coth{(\beta\omega'/2)}N^{zz}(\omega'),
\label{SCl}
\end{eqnarray}
 
respectively. Note that since SC is defined as a connected correlation function, the longitudinal contribution doesn't contain the term proportional to the equilibrium local moment. At T=0, SDF originate purely from the spin zero-point motion and we have $\left\langle\mathbf{s}^{2}\right\rangle^{t,l}_\omega=n^{t,l}(\omega)$. Therefore, the spin zero-point motion contribution to the SC is given by $n^{t,l}(\omega)$. SC is related to the effective fluctuating moment that is given by

\begin{equation}
m_{\text{eff}}(\omega)=\sqrt{\left(m_{\text{eff}}^{t}(\omega)\right)^2+\left(m_{\text{eff}}^{l}(\omega)\right)^2}.
\label{meff}
\end{equation}
 
Here, transverse and longitudinal contributions to $m_{\text{eff}}(\omega)$ are given by

\begin{equation}
m_{\text{eff}}^{t,l}(\omega)=\frac{g\mu_B}{\hbar}\sqrt{\left\langle\mathbf{s}^{2}\right\rangle^{t,l}_\omega},
\label{mtleff}
\end{equation}

where $g$ is the electron g-factor and $\mu_B$ is the Bohr magneton. Note that according to the above equations, the evaluation of the full ($\omega\rightarrow\infty$) values  of SC and the effective fluctuating moment involves integrals over all ranges of $\mathbf{q}$'s and $\omega$'s which makes such studies computationally demanding.

FDT can be also used to evaluate the value of the equilibrium local moment $m\equiv g\mu_B\int d\mathbf{r}\langle \hat{s}_z(\mathbf{r})\rangle$ (the spatial integration is over the atomic sphere). This leads to the following sum rule:

\begin{eqnarray}
m=m_{\text{a}}(\omega\rightarrow\infty),
\label{SumRuleEq}
\end{eqnarray}

where we defined an auxilary function 

\begin{equation}
m_{\text{a}}(\omega)=\frac{g\mu_B}{4\hbar^2}\int_0^\omega d\omega'\left[N^{+-}(\omega')-N^{-+}(\omega')\right].
\label{m0omega}
\end{equation}

\subsection{Dynamic susceptibility calculations}

The dynamic spin susceptibility tensor was evaluated using the linear response time-dependent density functional theory (TDDFT) within the local spin density approximation (LSDA).\cite{CW,Gross} This technique has been employed for dynamic spin susceptibility calculations in a number of systems.\cite{Savrasov,Staunton,Costa,Buczek,Bergara,Lounis} In this formalism, one first considers the Kohn-Sham (bare) susceptibility function given by

\begin{eqnarray}
\nonumber
\chi_0^{\alpha\beta}(\mathbf{r},\mathbf{r}',\mathbf{q},\omega)=\sum_{\mathbf{k}}^{\text{BZ}}\sum_{n,m}\sum_{\sigma\sigma'}\left(f^{\sigma}_{n\mathbf{k}}-f^{\sigma'}_{m\mathbf{k}+\mathbf{q}}\right) \\
\times\sigma_{\sigma\sigma'}^{\alpha}\sigma_{\sigma'\sigma}^{\beta}\frac{\psi^{\sigma*}_{n\mathbf{k}}(\mathbf{r})\psi^{\sigma'}_{m\mathbf{k}+\mathbf{q}}(\mathbf{r})\psi^{\sigma'*}_{m\mathbf{k}+\mathbf{q}}(\mathbf{r}')\psi^{\sigma}_{n\mathbf{k}}(\mathbf{r}')}{\hbar\omega+\epsilon^\sigma_{n\mathbf{k}}-\epsilon^{\sigma'}_{m\mathbf{k}+\mathbf{q}}+i\hbar\eta},
\label{chi0}
\end{eqnarray}

where $\sigma_{\sigma\sigma'}^{0}=\delta_{\sigma\sigma'}$ and $\sigma_{\sigma\sigma'}^{x,y,z}$ are elements of the Pauli matrices. The Kohn-Sham eigenfunctions $\psi^\sigma_{n\mathbf{k}}$ and eigenenergies $\epsilon^{\sigma}_{n\mathbf{k}}$ (the $n$, $\mathbf{k}$, and $\sigma$ indices denote band, wavevector, and spin quantum number, respectively), are obtained from standard LSDA calculations. Finally, $f^\sigma_{n\mathbf{k}}\equiv f(\epsilon^\sigma_{n\mathbf{k}})$ is the Fermi-Dirac distribution function. 

The (enhanced) susceptibility is then given by the Dyson-like equation

\begin{eqnarray}
\nonumber
\chi^{\alpha\beta}(\mathbf{r},\mathbf{r}',\mathbf{q},\omega)= \chi_0^{\alpha\beta}(\mathbf{r},\mathbf{r}',\mathbf{q},\omega) + \sum_{\gamma\delta}\int d\mathbf{r}_1d\mathbf{r}_2 \\
\times\chi_0^{\alpha\gamma}(\mathbf{r},\mathbf{r}_1,\mathbf{q},\omega)f_{\text{Hxc}}^{\gamma\delta}(\mathbf{r}_1,\mathbf{r}_2,\mathbf{q})\chi^{\delta\beta}(\mathbf{r}_2,\mathbf{r}',\mathbf{q},\omega).
\label{Dyson}
\end{eqnarray}

Here,

\begin{eqnarray}
\nonumber
f^{\alpha\beta}_{\text{Hxc}}(\mathbf{r},\mathbf{r}',\mathbf{q})=e^2\delta_{\alpha 0}\delta_{\beta 0}\sum_{\mathbf{R}}\frac{\text{exp}(-i\mathbf{q}\cdot\mathbf{R})}{|\mathbf{R}+\mathbf{r}-\mathbf{r}'|}\\
+f^{\alpha\beta}_{\text{xc}}(\mathbf{r})\delta(\mathbf{r}-\mathbf{r'}),
\label{fhxc}
\end{eqnarray}

where $f^{\alpha\beta}_{\text{xc}}(\mathbf{r})$ is the adiabatic local density approximation to the exchange-correlation kernel.\cite{Gross} For numerical calculations, some finite basis must be chosen to represent the spatial dependence of $\chi^{\alpha\beta}$, $\chi_0^{\alpha\beta}$, and $^{\alpha\beta}_{\text{Hxc}}$ functions. Eq. (\ref{Dyson}) can be then solved by matrix inversion. The quantities defined in the previous section can be subsequently evaluated using both $\chi^{\alpha\beta}$ and $\chi_0^{\alpha\beta}$. In the latter case, we refer to them as 'bare' quantities and denote them by using subscript $0$. 

From the computational point of view, the convergence with respect to the basis size as well as an accurate evaluation of the bare susceptibility at high energies are major challenges. For this reason the calculations were performed using two independent computational techniques (see below). In addition, we ensured reliability of the results by checking the sum rule in Eq. (\ref{SumRuleEq}). Note that both $\chi_0^{\alpha\beta}$ and $\chi^{\alpha\beta}$ satisfy the sum rule with the same LSDA local moment.\cite{Edwards} This allows us to independently gauge accuracy of both $\chi_0^{\alpha\beta}$ and $\chi^{\alpha\beta}$ calculations.

First computational method (below as Method I) is based on the real space finite temperature Matsubara technique. In this formalism, one does not evaluate $\chi_0^{\alpha\beta}(\mathbf{r},\mathbf{r}',\mathbf{q},\omega)$ from Eq. (\ref{chi0}) since it is computationally demanding due to presence of the summation over unoccupied states that is entangled with the BZ summation. Instead, one considers the Kohn-Sham susceptibility in the Matsubara time domain. This function can be efficiently evaluated in the real space according to

\begin{eqnarray}
\nonumber
\chi_{0}^{\alpha\beta}(\mathbf{r},\mathbf{r}',\mathbf{q},\tau)=\sum_{\mathbf{R}}e^{-i\mathbf{q}\cdot\mathbf{R}}\sum_{\sigma\sigma'}\sigma_{\sigma\sigma'}^{\alpha}\sigma_{\sigma'\sigma}^{\beta} \\
G^{\sigma}_{\mathbf{R}}(\mathbf{r},\mathbf{r}',\tau)G^{\sigma'}_{-\mathbf{R}}(\mathbf{r}',\mathbf{r},\beta-\tau).
\end{eqnarray}

Here, $\tau$ is the Matsubara time ($0\leq\tau\leq\beta$) and $G^{\sigma}_{\mathbf{R}}(\mathbf{r},\mathbf{r}',\tau)$ is the imaginary-time Kohn-Sham Green's function given by

\begin{equation}
G^{\sigma}_{\mathbf{R}}(\mathbf{r},\mathbf{r}',\tau)=-\sum_{n\mathbf{k}}f^{\sigma}_{n\mathbf{k}}\psi^\sigma_{n\mathbf{k}}(\mathbf{r})\psi^{\sigma*}_{n\mathbf{k}}(\mathbf{r}')e^{i\mathbf{k}\cdot\mathbf{R}}e^{\epsilon^{\sigma}_{n\mathbf{k}}\tau/\hbar}.
\label{gf}
\end{equation}

Then, the Kohn-Sham susceptibility is transformed into the Matsubara frequency domain [$\chi_{0}^{\alpha\beta}(\mathbf{r},\mathbf{r}',\mathbf{q},i\omega_k)$ with $\omega_k=\frac{2\pi k}{\hbar\beta}$ being a bosonic Matsubara frequency and $k$ being an integer] according to the prescription from Ref. \onlinecite{Kutepov2}. Subsequently, the enhanced susceptibility in Matsubara frequency domain was found from Eq. (\ref{Dyson}).

The calculations were based on the full-potential linear augmented plane waves (FLAPW) method as implemented in our in-house electronic structure code.\cite{Kutepov} The spatial dependence of the susceptibility functions is represented using the mixed product basis set that consists of numerical functions inside the muffin-tin spheres and plane/dual-plane waves in the interstitial region.\cite{PB} The specific expressions for $\chi_{0}^{\alpha\beta}(\mathbf{r},\mathbf{r}',\mathbf{q},\tau)$ in the product basis are analogous to those used for calculations  of the polarizability in Ref. \onlinecite{Kutepov2}. 

The Matsubara time real space formalism allows for very efficient susceptibility calculations. In addition, the frequency integrals up to infinity [e.g., Eqs. (\ref{SCt}) and (\ref{SCl})] can be very accurately evaluated on the imaginary frequency axis (see Ref. \onlinecite{Kutepov2}). The real frequency axis (with a small imaginary part $\eta$=1 meV) results need to be obtained by analytical continuation.  We employ an analytical continuation based on the continued fraction expansion method.\cite{Vidberg} It is designed to obtain an accurate representation of the low-energy spectrum but may become unstable at higher energies. Therefore, it is important to check the results of our calculations against an alternative approach.

For this reason the most important results were recalculated using the second method (below as Method II). In this approach the susceptibility is found using the technique implemented in the FLAPW \textsc{elk} code.\cite{elk} Here, $\chi_0^{\alpha\beta}(\mathbf{r},\mathbf{r}',\mathbf{q},\omega)$  is evaluated directly from Eq. (\ref{chi0}) and the spatial dependence of the susceptibility functions is represented using the plane wave basis. Since it works on the real frequency axis (with a small imaginary part), Method II does not involve analytical continuation. However, it is difficult to converge the results especially at high energies. In addition, a lot of plane waves are needed to obtain an accurate description of the spatial dependence. Consequently, Method II is significantly more computationally expensive than Method I.

\subsection{Computational Details}

We consider Fe (bcc), Co (fcc), and Ni (fcc) 3\emph{d} ferromagnets with experimental lattice parameters. A 16$\times$16$\times$16 k-point mesh was used. For the FLAPW basis the energy cutoff in the interstitial region was set to at least 12 Ry and the angular momentum cutoff inside the muffin-tin sphere was set to $L_{\text{max}}=8$.  In addition, the local orbitals were included in order to ensure an accurate description of the excited states which is crucial for SDF studies. We found that inclusion of local orbitals for the 3\emph{s}, 3\emph{p}, and 4\emph{d} states was sufficient to obtain converged results. 

For Method I, $T$=300 K and we used 158 nonuniformly distributed (see Ref. \onlinecite{Kutepov2} for details) mesh points on the imaginary Matsubara time axis. The mixed product basis set was constructed using the interstitial energy cutoff 16.5 Ry and the muffin-tin angular momentum cutoff $L_{\text{max}}^{\text{PB}}=4$. 

For Method II, the $G$ vector cutoff for the plane wave basis was set to 9.6 $\AA^{-1}$. For the bare susceptibility calculations, all unoccupied states up to 3.2 Ry above the Fermi energy were included. 

For both methods, we ensured that the results are well converged with respect to the computational parameters.

\section{Results and Discussion}

\subsection{Small wavevector SDF}

\begin{figure}[t!]
\includegraphics[width=1.0\hsize]{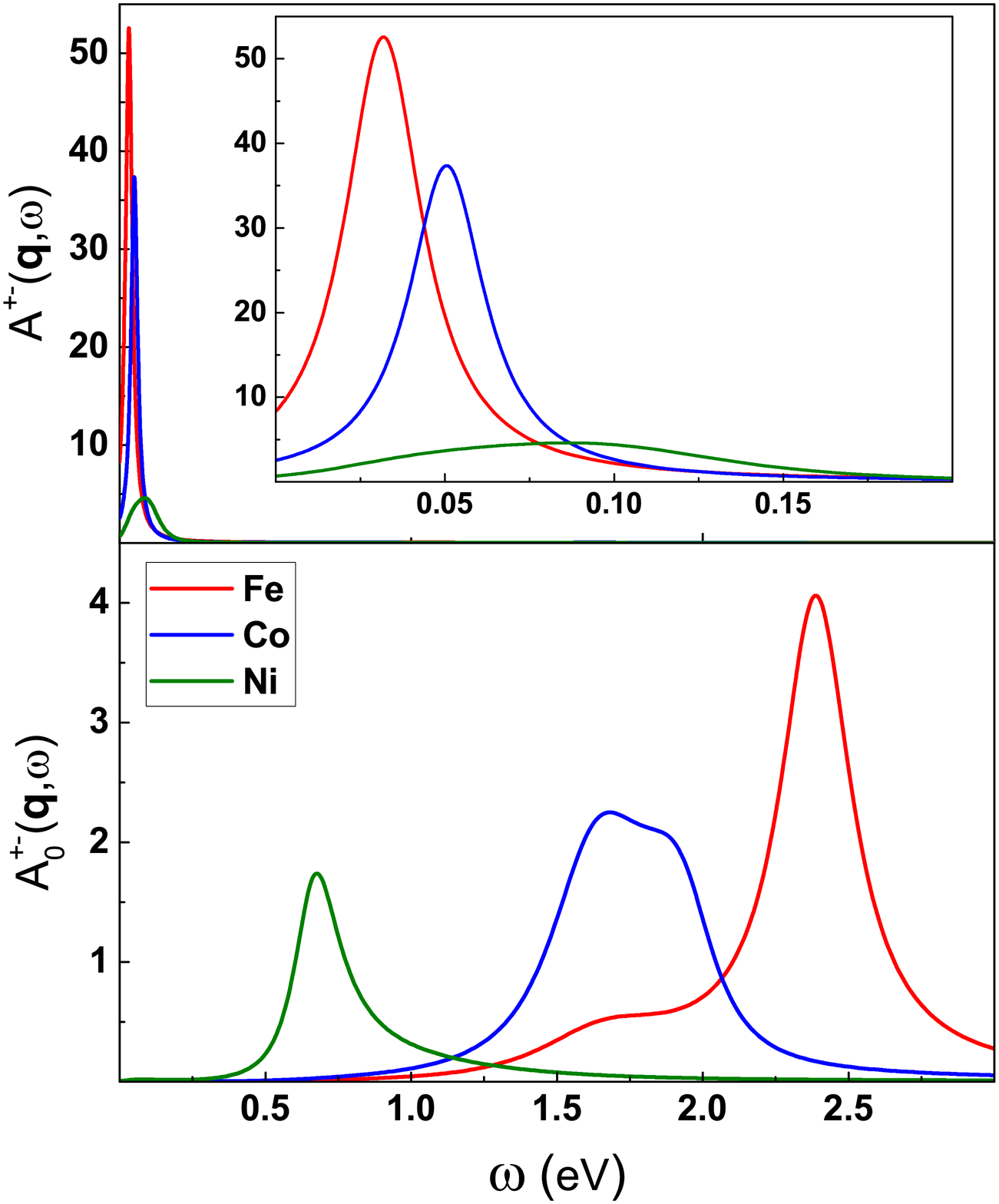}
\caption{Small wavevector transverse SDF for Fe, Co, and Ni calculated using Method I. Top: transverse spectral function. The inset shows the low-energy part of the plot. Bottom: 'bare' transverse spectral function. We used $\mathbf{q}=(0,0,0.125)2\pi/a$. Vertical axis units are $\hbar^2/\text{eV}$. Well-defined spin wave excitations exist at low energies.}
\label{ApmLowQ}
\end{figure}

Let us first consider SDF for small wavevectors. Fig. \ref{ApmLowQ} shows the transverse spectral function (top) and the 'bare' spectral function (bottom) for a fixed low magnitude $\mathbf{q}$ as a function of the frequency for Fe, Co, and Ni calculated using Method I. For all materials, $A^{+-}(\mathbf{q},\omega)$ has a well defined peak at low energies (below 0.1 eV). As we increase the number of 3\emph{d} electrons moving from Fe to Co and then to Ni, the peak moves to higher energies, its amplitude decreases, and its  width increases. This is in agreement with previous studies.\cite{Buczek} At higher energies (above 0.5 eV) $A^{+-}(\mathbf{q},\omega)$ is very small. On the other hand, $A_0^{+-}(\mathbf{q},\omega)$ is negligible at low energies but it has a nontrivial structure at higher energies. In particular, we observe a broad maximum at around 0.75 eV, 1.75 eV, and 2.5 eV for Ni, Co, and Fe, respectively. This maximum originates from single-particle Stoner excitations that correspond to spin-flip electronic transitions between majority and minority bands. Our results indicate that many-body correlations suppress these high-energy Stoner excitations and instead produce low-energy collective spin wave modes that are responsible for the $A^{+-}(\mathbf{q},\omega)$ peaks. The nonzero width of the peaks indicates a finite lifetime of the spin waves due to interaction with Stoner excitations (Landau damping). Indeed, while it is not explicitly seen in the figure, the $A_0^{+-}(\mathbf{q},\omega)$ weight in the low energy region increases with the number of 3\emph{d} electrons and leads to the corresponding increase of the width of the spin wave peaks.

An important feature of ferromagnetic systems in the absence of external magnetic field and spin-orbit coupling is the presence of a uniform ($\mathbf{q}$=0) zero frequency Goldstone mode. It is well known, however, that numerical calculations based linear response TDDFT method produce spurious finite frequency of the Goldstone mode due to inconsistency between the calculations of the exchange-correlation kernel and the Kohn-Sham susceptibility.\cite{Buczek,Bergara,Lounis3} In particular, our calculations yield the Goldstone mode frequency of 10-40 meV and, consequently, the energies of low-$\mathbf{q}$ excitations (Fig. \ref{ApmLowQ} top) are overestimated. In order to cure this problem, several correction schemes have been designed based on a modification of the exchange-correlation kernel\cite{Buczek,Lounis3} or Kohn-Sham susceptibility\cite{Bergara} such that the zero-frequency Goldstone mode is recovered. While such a procedure is crucial for spin wave dispersion studies, in this work we focus on BZ-integrated quantities at much larger energy scales and, therefore, the presence of finite excitation gap of the order of few tens meV has a small effect on these results.

\begin{figure}[t!]
\includegraphics[width=1.0\hsize]{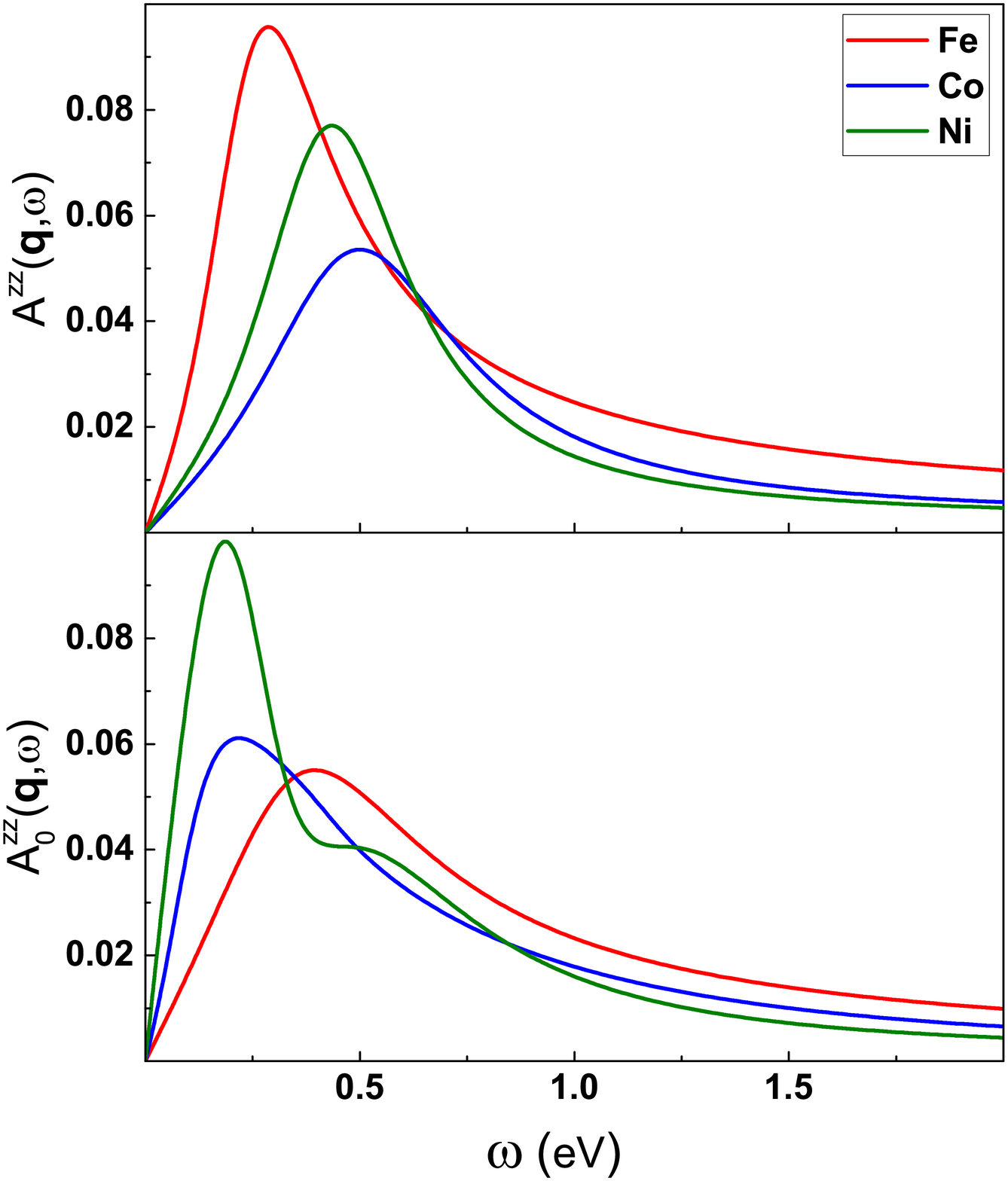}
\caption{Small wavevector longitudinal SDF for Fe, Co, and Ni calculated using Method I. Top: longitudinal spectral function. Bottom: 'bare' longitudinal spectral function. We used $\mathbf{q}=(0,0,0.125)2\pi/a$. Vertical axis units are $\hbar^2/\text{eV}$. Low-$\mathbf{q}$ longitudinal SDF are significantly smaller than the transverse one.}
\label{AzzLowQ}
\end{figure}

The low $\mathbf{q}$ longitudinal spectral functions calculated using Method I is shown in Fig. \ref{AzzLowQ}. For all materials $A^{zz}(\mathbf{q},\omega)$ ( Fig. \ref{AzzLowQ} top) has a broad peak structure that slowly decays with energy. The 'bare'  longitudinal spectral function (Fig. \ref{AzzLowQ} bottom) has the majority of weight in the same energy range as $A^{zz}(\mathbf{q},\omega)$ with only a slightly lower amplitude. Since $A_0^{zz}(\mathbf{q},\omega)$ describes electronic transitions within the same spin channel, we can conclude that the low $\mathbf{q}$ longitudinal SDF originate predominantly from the spin-conserving single-particle excitations. However, the overall magnitude of $A^{zz}(\mathbf{q},\omega)$ is  substantially smaller from $A^{+-}(\mathbf{q},\omega)$. This indicates that for small $\mathbf{q}$ values the longitudinal SDF can be neglected and only transverse SDF play an important role.

\subsection{Density of SDF}

\begin{figure}[t!]
\includegraphics[width=1.0\hsize]{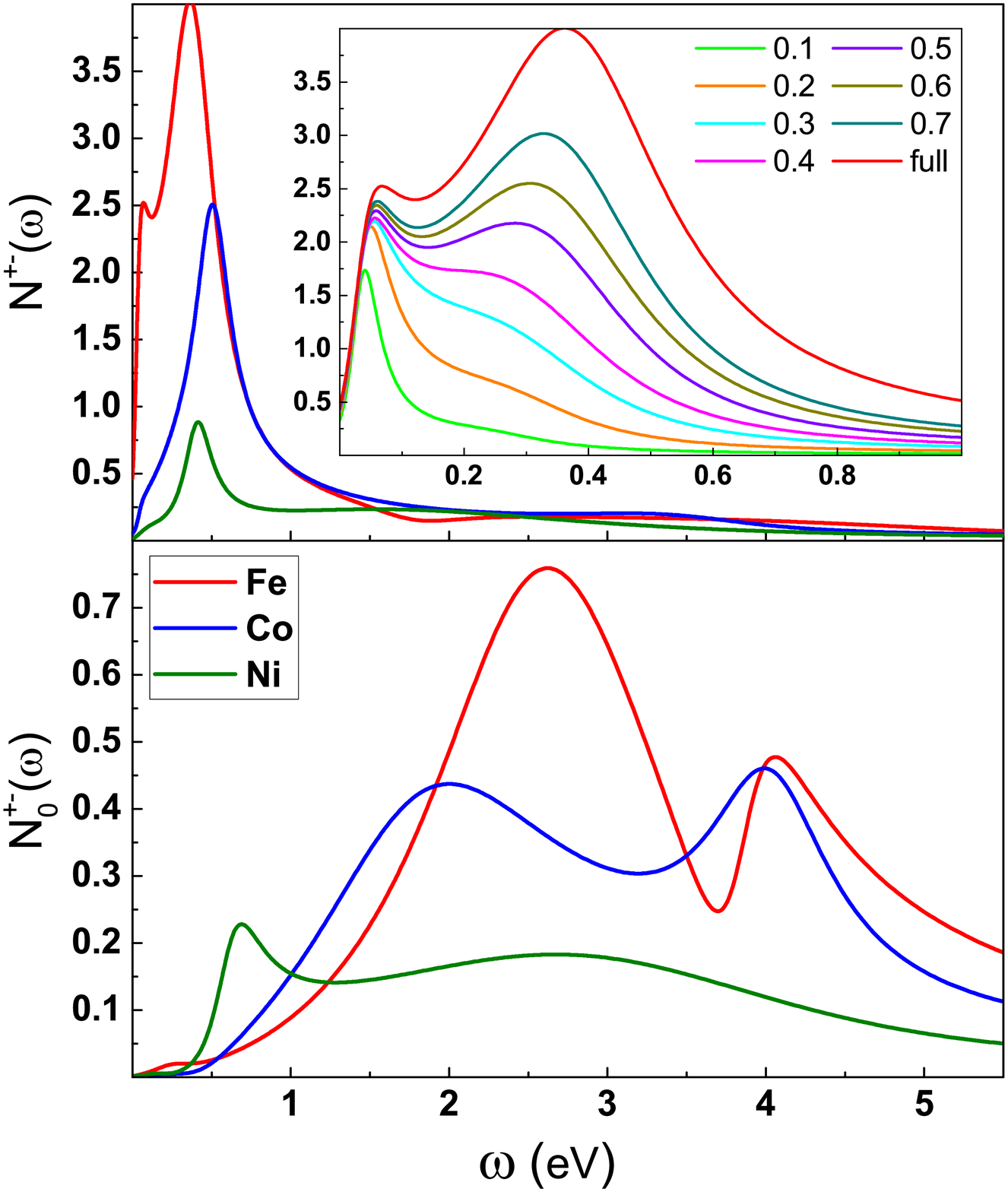}
\caption{On-site transverse SDF spectrum for Fe, Co, Ni calculated using Method I. Top: density of transverse SDF. The inset shows the low energy part of the plot for Fe (the red curve denoted as 'full') compared with the partial-$\mathbf{q}$ density of transverse SDF [see Eq. (\ref{NrSDF}), different curves are denoted by the value of the $\Omega_{\mathbf{q}}/\Omega_{\text{BZ}}$ ratio]. Bottom: 'bare' density of transverse SDF. Vertical axis units are $\hbar^2/\text{eV}$. Transverse SDF in 3\emph{d} ferromagnets show a generic two-peak structure.}
\label{Npm}
\end{figure} 

\begin{figure}[t!]
\includegraphics[width=1.0\hsize]{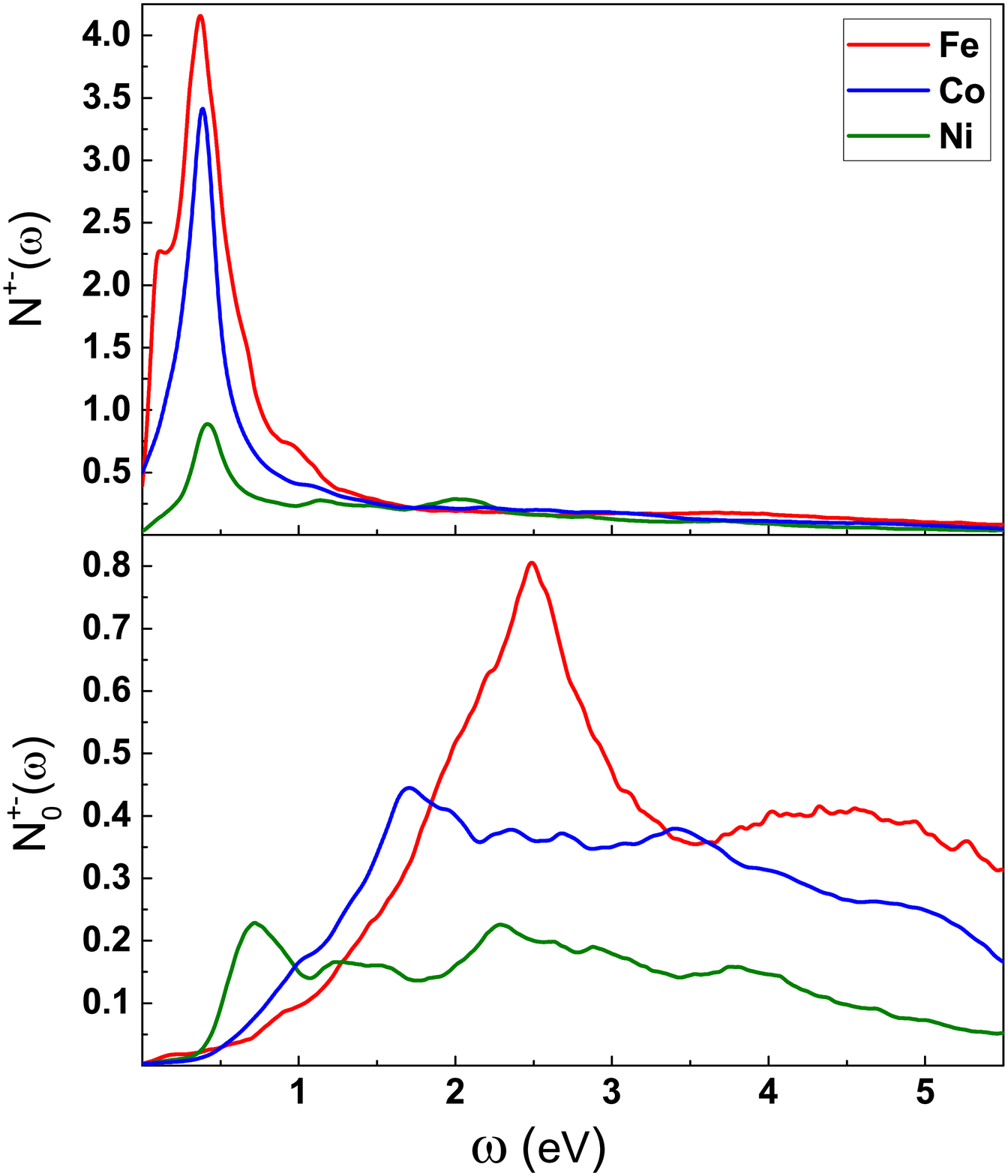}
\caption{On-site transverse SDF spectrum for Fe, Co, Ni calculated using Method II. Top: density of transverse SDF. Bottom: 'bare' density of transverse SDF. Vertical axis units are $\hbar^2/\text{eV}$. Spectra calculated using method II are in a good agreement with the results obtained using Method I.}
\label{NpmElk}
\end{figure}

Let us now analyze the density of SDF.  According to Eq. (\ref{NSDF}), this function includes SDF from the entire BZ. We focus on transverse SDF. Fig. \ref{Npm} shows $N^{+-}(\omega)$ and $N^{+-}_0(\omega)$ for all considered materials calculated using Method I. The same quantities but calculated using Method II are shown in Fig. \ref{NpmElk}. For all materials, both methods produce similar $N^{+-}(\omega)$ curves although some differences in linewidths can be observed. In the case of $N^{+-}_0(\omega)$ overall, we also have a reasonable agreement except for $\omega>$ 3.5 eV where we have some deviation. This is the energy region where the adopted analytical continuation procedure may be inaccurate.

We find that most of the $N^{+-}(\omega)$ weight exists for $\omega<$ 1 eV. On the other hand, $N^{+-}_0(\omega)$ (that describes the spectrum of single-particle Stoner excitations) is much smaller in this energy range but instead it extends to much higher energies with the majority of the spectrum residing up to an energy of the order of the 3\emph{d} electronic bandwidth ($W_{\text{el}} \simeq$ 5-6 eV). Therefore, similarly as in the case of small $\mathbf{q}$ SDF, we conclude that many-body interactions suppress the high-energy Stoner excitations and transform them into low-energy collective modes.

For Fe, $N^{+-}(\omega)$ has a two-peak structure with the smaller narrow low-energy peak at 50 meV and the larger broad high-energy peak at 0.4 eV. While for Co and Ni only the high-energy peak can be clearly seen, for both materials we can also identify a low-energy shoulder at $\sim$ 50 meV. This indicates that the two-peak structure is a generic feature for the 3\emph{d} magnets. We emphasize that the shape of $N^{+-}(\omega)$ is, thus, distinctly different from the single-peak structure of the spectral function. This indicates that transverse excitations with large wavevectors play an important role. This point is quantitatively illustrated in the inset of Fig. \ref{Npm} in the case of Fe. Here, the partial-$\mathbf{q}$ density of transverse SDF, Eq. (\ref{NrSDF}), is shown for different values of the $\Omega_{\mathbf{q}}/\Omega_{\text{BZ}}$ ratio. As seen, for $\omega<$ 0.1 eV, SDF with a small $\mathbf{q}$ that correspond to traditional spin wave excitations are dominant and they are responsible for the low-energy peak. For higher energies, however, SDF with a large $\mathbf{q}$ are crucial. In particular, the large high-energy peak originates exclusively from collective excitations with large $\mathbf{q}$ values that are localized in the real space. Analysis of $N_{\Omega_\mathbf{q}}^{+-}(\omega)$ for Co and Ni shows that the origin of the two-peak structure is similar for all considered systems.

The above discussion indicates that in order to properly include SDF in calculations of ground state and thermodynamic properties, one needs to take into account excitations for all $\mathbf{q}$. Therefore, restriction to SDF from only limited parts of the BZ (for instance the long wave approximation commonly used in spin fluctuation theories or the DMFT single-site approximation) can lead to an inaccurate material description and misleading results.

\subsection{Local moment sum rule}

\begin{figure}[t!]
\includegraphics[width=1.0\hsize]{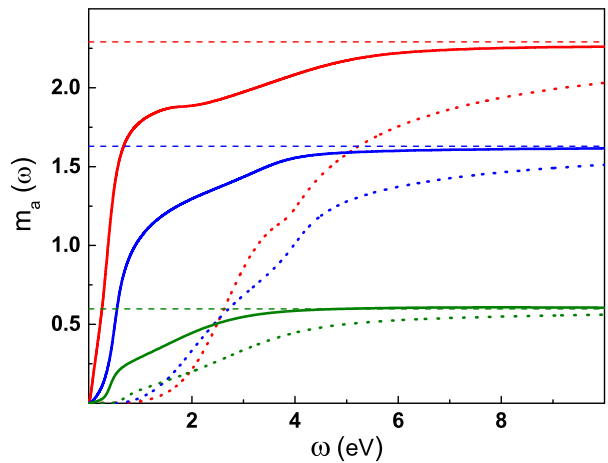}
\caption{Local moment sum rule (see Eq. (\ref{SumRule})) evaluated using Method I. Red, blue, and green curves correspond to Fe, Co and Ni, respectively. Horizontal dashed lines denote the LSDA value of the local moment. Full and dotted line denote $m_{\text{a}}(\omega)$ and $m_{\text{a},0}(\omega)$, respectively. Vertical axis units are $\mu_B$. For all materials the sum rule is satisfied by including SDF up to energy of the order of $W_{\text{el}}$.}
\label{SumRule}
\end{figure}

\begin{figure}[t!]
\includegraphics[width=1.0\hsize]{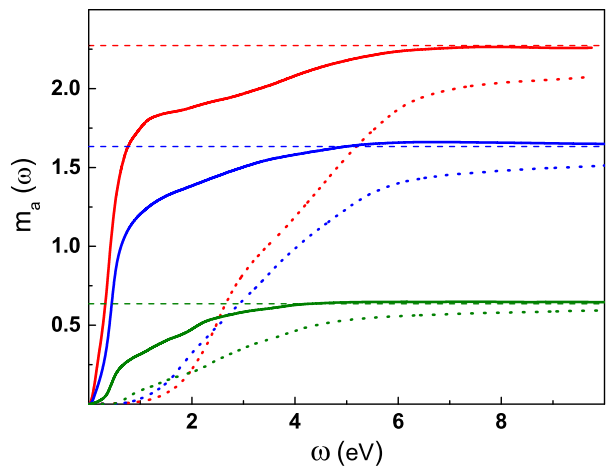}
\caption{Local moment sum rule (see Eq. (\ref{SumRule})) evaluated using Method II. Red, blue, and green curves correspond to Fe, Co and Ni, respectively. Horizontal dashed lines denote the LSDA value of the local moment. Full and dotted line denote $m_{\text{a}}(\omega)$ and $m_{\text{a},0}(\omega)$, respectively. Vertical axis units are $\mu_B$. Method II produce results similar to Method I even at high energies.}
\label{SumRuleElk}
\end{figure}

In this section we analyze the local moment sum rule in Eq. (\ref{SumRuleEq}).  Fig. \ref{SumRule} shows $m_{\text{a}}(\omega)$ for Fe, Co, and Ni evaluated from both $\chi^{\alpha\beta}$ and $\chi_0^{\alpha\beta}$. The LSDA values of the local moment are shown as dashed horizontal lines. The same plot but obtained using Method II is shown in Fig. \ref{SumRuleElk}. As seen, the sum rule is almost perfectly satisfied in both sets of calculations.  The shapes of the $m_{\text{a}}(\omega)$ curves are also very similar in both methods (even at high energies). This is especially true for the enhanced susceptibility. These results demonstrate that our calculations maintain high level of accuracy up to very high energies. In particular, we can conclude that different independent basis sets used in both methods are well converged and analytical continuation is quite reliable.
 
Note that for Fe, $m_{\text{a}}(\omega)$ becomes close to the LSDA local moment value already at the energies of the order of $W_{\text{el}}$. On the other hand, for the 'bare' SDF spectrum, energies up to 13 eV are required to obtain a similar level of accuracy. In the case of system with smaller moment (like Ni and Co) such convergence is obtained for lower energies.

\subsection{Number of SDF}

\begin{figure}[t!]
\includegraphics[width=1.0\hsize]{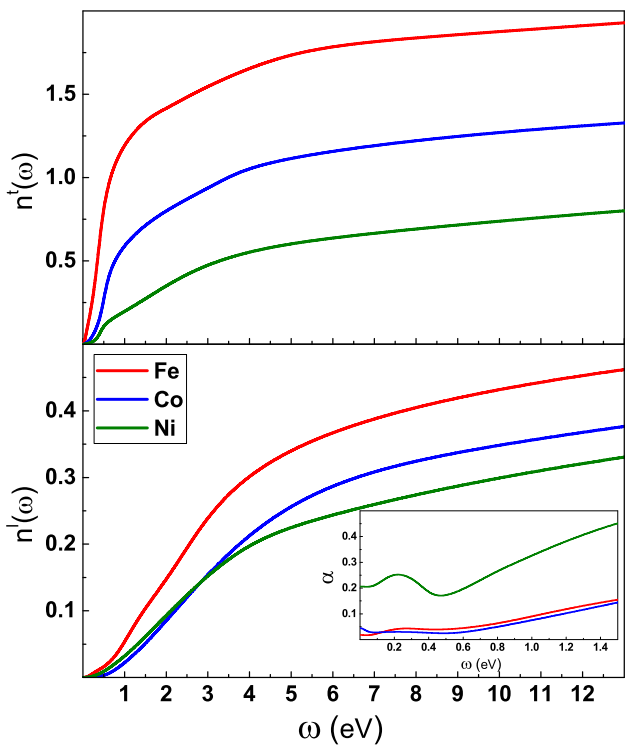}
\caption{Energy dependence of the number of on-site transverse SDF for Fe, Co, and Ni calculated using Method I. Top: number of on-site transverse SDF. Bottom: 'bare' number of on-site transverse SDF. Vertical axis units are $\hbar^2$. The inset shows the energy dependence of the adiabaticity parameter $\alpha$ defined as twice the ratio of $n^l(\omega)$ to $n^t(\omega)$.}
\label{nsf}
\end{figure} 

Let us now consider the number of SDF. The number of transverse SDF calculated using Method I is shown in Fig. \ref{nsf} as a function of energy for different ferromagnetic metals. As expected from the above analysis of $N^{+-}(\omega)$, the most of transverse SDF exist for $\omega<$ 1 eV with the high-energy peak providing the major contribution. Nevertheless, for $\omega>$ 1 eV, $n^t(\omega)$ still shows a sizable increase up to $\omega\sim W_{\text{el}}$. For $\omega>W_{\text{el}}$, only a slow increase of $n^t(\omega)$ is observed that corresponds to excitations involving semicore and/or high-energy unoccupied states.

The number of longitudinal SDF is shown in Fig. \ref{nsf} (bottom). As seen, longitudinal SDF exist at all energies with the majority of the spectrum being accumulated for $\omega<W_{\text{el}}$. While at low energies ($\omega<1$ eV) $n^l(\omega)<<n^t(\omega)$, for $\omega\sim W_{\text{el}}$ both functions have the same order of magnitude. Indeed, the longitudinal SDF do not disappear when local moments are present, but rather they are shifted to higher energies. Thus, our study naturally addresses the validity of the adiabatic approximation\cite{Antropov2} in spin dynamics which neglects the longitudinal spin dynamics. Quality of this approximation can be characterized by the adiabaticity parameter $\alpha$ defined as twice the ratio of $n^l(\omega)$ to $n^t(\omega)$. The energy dependence of this quantity is shown in the inset of Fig. \ref{nsf}. For Fe and Co, the adiabatic criterion\cite{Antropov2} is well fulfilled ($\alpha<0.1$ up to $\omega\sim 1$ eV) so pure transversal spin dynamics is valid in this energy region. We emphasize, however, that in our case of magnetic metals there is an important difference with a spin dynamics in magnetic insulators due to a presence of strong non spin wave transversal SDF of itinerant nature. In addition, for Ni $\alpha$ is significantly larger and the adiabatic criterion is not fulfilled so the itinerant longitudinal SDF play an important role in spin dynamics.

We emphasize that for both transverse and longitudinal SDF, the majority of excitations lie at energies much higher than those accessible from inelastic neutron scattering experiments. Therefore, different experimental techniques (high-energy spin resolved spectroscopies\cite{exp}) are required to probe the full spectrum. Both $n^t(\omega)$ and $n^l(\omega)$ are continuous steadily increasing functions of energy and therefore, it is not possible to rigorously introduce any energy cutoff when including SDF in studies of metals. Thus, with a temperature increase for instance, more SDF are excited and contribute to the magnetic properties of the itinerant metal. This feature is in stark contrast with the traditional magnetic insulator picture where excitations for energies above the spin wave spectrum do not exist and \emph{all} SDF are excited at corresponding temperatures.

\subsection{FDT}

\begin{table}
\caption{Effective fluctuating moment ($\mu_B$) calculated using Method I at different energies for all considered materials. Note that $m_{\text{eff}}(\omega)$ does not contain contribution from the equilibrium local moment. The zero values correspond to the calculated values that are less than 0.1 $\mu_B$.} 
\begin{ruledtabular}
\begin{tabular}{l|ccccc}
$\omega$ (eV) & 0.1 & 1 & 5 & 12 & $\infty$ \\
\hline
\hbox{Fe:}    $m_{\text{eff}}^{t}(\omega)$   & 0.8 & 2.2 & 2.7 & 2.8 & 3.1 \\
\phantom{Fe:} $m_{\text{eff},0}^{t}(\omega)$ & 0.0 & 0.3 & 1.8 & 2.2 & 2.6 \\
\phantom{Fe:} $m_{\text{eff}}^{l}(\omega)$   & 0.1 & 0.5 & 1.2 & 1.3 & 1.6 \\
\phantom{Fe:} $m_{\text{eff},0}^{l}(\omega)$ & 0.0 & 0.2 & 1.1 & 1.2 & 1.5 \\
\phantom{Fe:} $m_{\text{eff}}(\omega)$       & 0.8 & 2.3 & 2.9 & 3.1 & 3.5 \\
\phantom{Fe:} $m_{\text{eff},0}(\omega)$     & 0.0 & 0.4 & 2.1 & 2.5 & 3.1 \\
\hline
\hbox{Co:}    $m_{\text{eff}}^{t}(\omega)$   & 0.2 & 1.5 & 2.1 & 2.3 & 2.7 \\
\phantom{Co:} $m_{\text{eff},0}^{t}(\omega)$ & 0.0 & 0.3 & 1.7 & 2.0 & 2.4 \\
\phantom{Co:} $m_{\text{eff}}^{l}(\omega)$   & 0.0 & 0.3 & 1.0 & 1.2 & 1.5 \\
\phantom{Co:} $m_{\text{eff},0}^{l}(\omega)$ & 0.1 & 0.3 & 1.0 & 1.2 & 1.5 \\
\phantom{Co:} $m_{\text{eff}}(\omega)$       & 0.2 & 1.6 & 2.3 & 2.6 & 3.1 \\
\phantom{Co:} $m_{\text{eff},0}(\omega)$     & 0.1 & 0.5 & 2.0 & 2.3 & 2.9 \\
\hline
\hbox{Ni:}    $m_{\text{eff}}^{t}(\omega)$   & 0.1 & 0.9 & 1.6 & 1.8 & 2.2 \\
\phantom{Ni:} $m_{\text{eff},0}^{t}(\omega)$ & 0.0 & 0.5 & 1.3 & 1.6 & 2.1 \\
\phantom{Ni:} $m_{\text{eff}}^{l}(\omega)$   & 0.0 & 0.4 & 0.9 & 1.1 & 1.5 \\
\phantom{Ni:} $m_{\text{eff},0}^{l}(\omega)$ & 0.1 & 0.3 & 0.9 & 1.1 & 1.4 \\
\phantom{Ni:} $m_{\text{eff}}(\omega)$       & 0.1 & 1.0 & 1.8 & 2.1 & 2.7 \\
\phantom{Ni:} $m_{\text{eff},0}(\omega)$     & 0.1 & 0.6 & 1.6 & 1.9 & 2.5 \\
\end{tabular}
\end{ruledtabular}
\label{tab1}
\end{table}

In this section we use FDT in order to evaluate SC and the related effective fluctuating moment. The calculations were made using Method I that allows for an efficient evaluation of the infinite energy integrals. 

Effective fluctuating moment $m_{\text{eff}}(\omega)$ provides a useful measure of the strength of SDF at a given energy since it can be compared with local moment values in magnetic materials. Note that $m_{\text{eff}}(\omega)$ is directly related to SC through Eqs. (\ref{meff}) and (\ref{mtleff}). Since the main contribution to SC arises from the spin zero-point motion SDF (except when $\omega<1/\beta$ where thermal SDF are important), the energy dependence of $m_{\text{eff}}^{t,l}(\omega)$ follows roughly the square root of $n^{t,l}(\omega)$. Therefore, $m_{\text{eff}}(\omega)$ is an ever increasing smooth function of energy. For this reason, it is sufficient to provide $m_{\text{eff}}(\omega)$ at several characteristic energy scales, see Table \ref{tab1}. Here, the values of $m_{\text{eff}}(\omega)$ as well as $m_{\text{eff}}^{t}(\omega)$ and $m_{\text{eff}}^{l}(\omega)$ calculated both from $\chi^{\alpha\beta}$ and $\chi_0^{\alpha\beta}$ using Method I are shown. At low energies ($\omega\simeq 0.1$ eV), $m_{\text{eff}}(\omega)$ originates mainly from traditional long-wavelength spin waves (low-energy peak in top panel of Fig. \ref{Npm}) and it is much smaller than $m$. For $\omega\simeq 1$ eV, the main part of the SDF spectrum that consists of localized in real space large $\mathbf{q}$ collective transverse excitations (high-energy peak in top panel of Fig. \ref{Npm}) is also included and $m_{\text{eff}}(\omega)$ becomes comparable to $m$. A further energy increase up to $\omega\simeq W_{\text{el}}$ includes all excitations within the 3\emph{d} band and $m_{\text{eff}}(\omega)$ is increased by 20-70\%. A large part of this enhancement originates from longitudinal SDF. For higher energies, only a slow increase of $m_{\text{eff}}(\omega)$ is observed. However, this accumulates to a significant contribution for $\omega=\infty$.

\begin{figure}[t!]
\includegraphics[width=1.0\hsize]{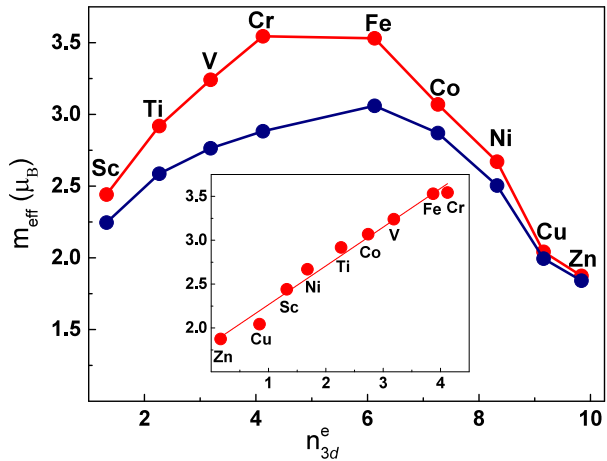}
\caption{The full effective fluctuating magnetic moment ($m_{\text{eff}}$) calculated using Method I as a function of the the number of 3\emph{d} electrons (red). Note that $m_{\text{eff}}$ does not contain contribution from the equilibrium local moment. The blue curve denotes the 'bare' $m_{\text{eff}}$ evaluated using the Kohn-Sham susceptibility. The inset shows $m_{\text{eff}}$ as a function of $n_{3d}=\text{min}\left(n^{e}_{3d},n^{h}_{3d}\right)$, where $n^{e}_{3d}$ and $n^{h}_{3d}$ is the number of 3\emph{d} electrons and holes, respectively. The line in the inset is the linear fit of the data. The effective fluctuating moment is independent on the presence of local moment and is determined solely by the 3\emph{d} band population.}
\label{Meff}
\end{figure}

In Fig. \ref{Meff} we plot $m_{\text{eff}}\equiv m_{\text{eff}}(\omega=\infty)$ and $m_{\text{eff},0}\equiv m_{\text{eff},0}(\omega=\infty)$ as a function of the number of 3\emph{d} electrons. In addition to the considered materials, we also included the data for 3\emph{d} paramagnetic metals from Ref. \onlinecite{Wysocki}. Interestingly, both $m_{\text{eff}}$ and $m_{\text{eff},0}$ seem not to be affected by the presence of local moments, but they are rather determined by the 3\emph{d} band population. Indeed, the dependence of  both quantities on the 3\emph{d} electron number is reminiscent of the Slater Pauling curve. Below the half-filling, their values increase with the 3\emph{d} electron number. Above the half-filling, an opposite trend is observed. This behavior follows from the well-known universal dependence of the imaginary part of a 'bare' response function on the electronic population which shows maximum for the Fermi level in the middle of the band. The enhanced susceptibility shows the same qualitative structure with additional enhancement that is the strongest close to half-filling. Note that similar curve was obtained for magnetic adatoms on metallic surfaces.\cite{Lounis2}

In the inset of Fig. \ref{Meff}, we show $m_{\text{eff}}$ as a function of the number of 3\emph{d} carriers as $n_{3d}=\text{min}\left(n_{3d}^{e},n_{3d}^{h}\right)$. Here, $n_{3d}^{e}$ and $n_{3d}^{h}$ is the number of 3\emph{d} electrons and holes, respectively. We find that $m_{\text{eff}}$ shows approximately a linear dependence on $n_{3d}$. The fitting to a linear function results in the following empirical formula: 

\begin{equation}
m_{\text{eff}} \approx 0.4 n_{3d}  +1.8.
\end{equation} 

Note that the same equation was obtained in Ref. \onlinecite{Wysocki} using only 3\emph{d} paramagnets. This indicates that every 3\emph{d} electron or hole contributes approximately the moment of 0.4$\mu_{B}$ to $m_{\text{eff}}$. The nonzero intercept corresponds to $m_{\text{eff}}$ for a completely filled or completely empty 3\emph{d} band. It originates from electronic transitions involving semicore levels and high-energy unoccupied states. We are not familiar with any theoretical or experimental discussion of such large contribution from semicore and high-energy states.

\begin{figure}[t!]
\includegraphics[width=1.0\hsize]{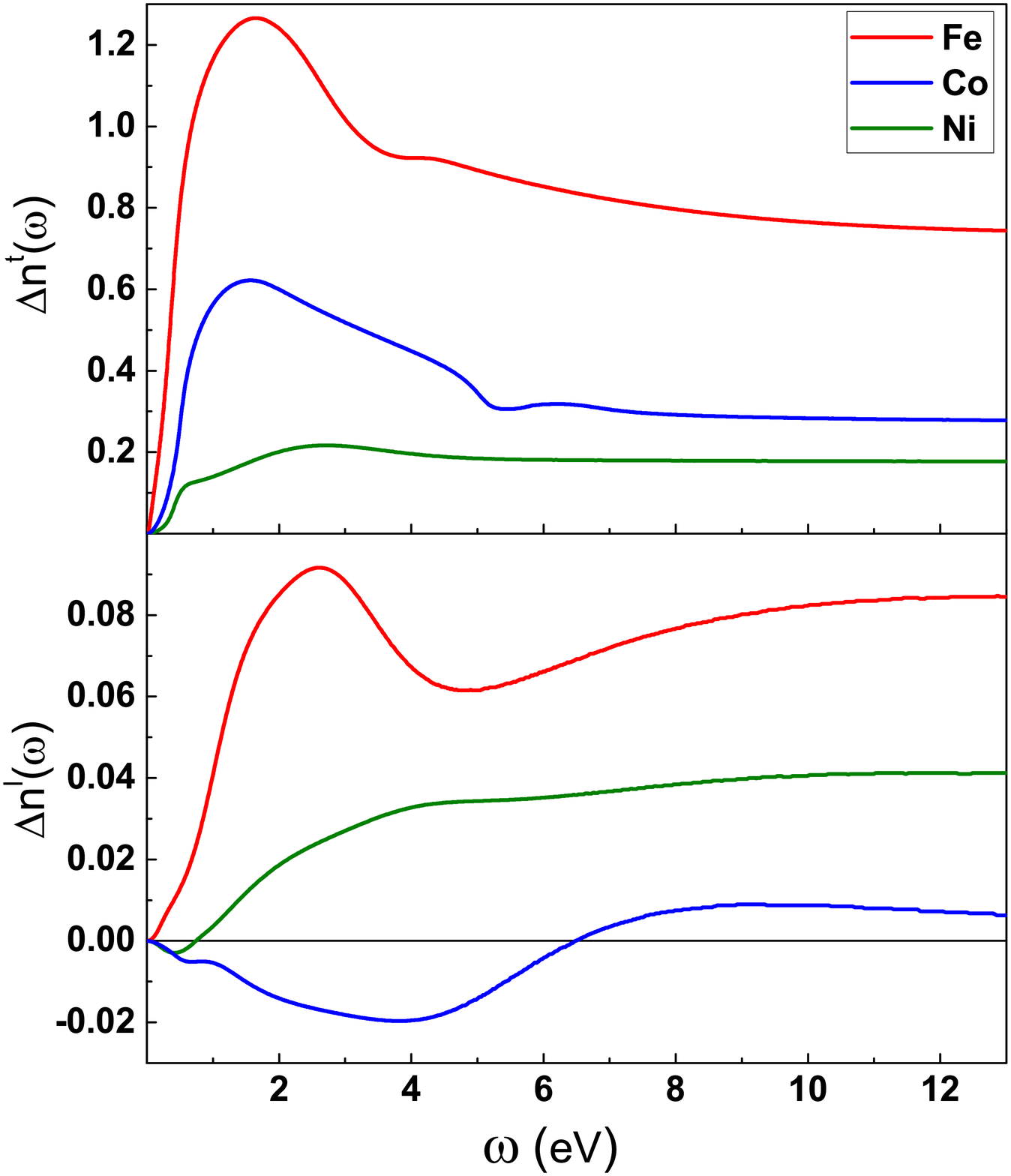}
\caption{Difference between the number and the 'bare' number of transverse (top) and longitudinal (bottom) SDF calculated using Method I. Vertical axis units are $\hbar^2$. Transverse SDF up to energy of the order of $W_{\text{el}}$ should be explicitly included in electronic structure calculations.}
\label{Delnsf}
\end{figure}

While $m_{\text{eff}}$ is a useful quantity that characterizes the overall strength of SDF, it is the difference between $m^2_{\text{eff}}$ and $m^2_{\text{eff},0}$ that determines the corresponding correlation energy (see, for instance, the recent review Ref. \onlinecite{RPA}). Indeed, the SDF correlation energy can be roughly estimated as a magnetic energy required to form the moment $\Delta m=\sqrt{m^2_{\text{eff}}-m^2_{\text{eff},0}}$. It follows then from Fig. \ref{Meff} that the SDF correlation energy is the largest close to the half-filling where the many-body enhancement is the strongest. In particular, $\Delta m$ is equal to 1.7 $\mu_B$, 1.1 $\mu_B$, and 0.9 $\mu_B$ for Fe, Co, and Ni, respectively. Clearly, $\Delta m$ is comparable to $m$ for all 3\emph{d} ferromagnets and, therefore, SDF should be included in electronic structure calculations for these materials. For Ni SDF are expected to be particularly important since the $\Delta m/m$ ratio is roughly twice as large as for Fe or Co. Note that for early 3\emph{d} paramagnets SDF should have even stronger effect on materials properties since the local moment is zero.\cite{Wysocki} In order to understand the energy distribution of SDF that contribute to the correlation energy, in Fig. \ref{Delnsf} we plotted the energy dependence of $\Delta n^{t,l}(\omega)=n^{t,l}(\omega)-n^{t,l}_0(\omega)$ (this quantity correspond to the $\Delta m^2$ at $T=0$). As seen, for all 3\emph{d} ferromagnets $\Delta n^t(\omega)$ converges for $\omega\sim W_{\text{el}}$ and, therefore, all SDF up to this energy should be included on equal footing in electronic structure calculations of these materials. Note that $\Delta n^l(\omega)<<\Delta n^t(\omega)$ so the contribution of longitudinal SDF to the correlations energy can be neglected.

\section{Conclusions}

SDF in 3\emph{d} ferromagnetic metals were analyzed for all spatial and time scales using first principles electronic structure calculations of the dynamic spin susceptibility tensor. The accuracy of the results were carefully tested by using two independent calculation methods and ensuring that the local moment sum rule is satisfied both for enhanced and bare susceptibilities.

We demonstrated that the SDF are spread continuously over the entire BZ as well as the wide energy range extending far above the 3\emph{d} bandwidth. Thus, no well-defined wavevector and frequency cutoffs (as often assumed) can be reliably introduced in such materials. Since the majority of excitations lie at energies much higher than those accessible by inelastic neutron scattering measurements, different experimental techniques, like spin-polarized high-energy spectroscopies, are required to probe the full SDF spectrum.

It was shown that the on-site SDF spectrum of 3\emph{d} ferromagnets has a generic structure that consists of two main constituents. One, at low energies (for instance, for Fe at $\sim$50 meV) is a minor contribution due to traditional low-$\mathbf{q}$ spin wave excitations, while the second, much larger high-energy (for instance, for Fe at $\sim$0.4 eV) component, corresponds to localized in real space large wavevector spin excitations. In addition, our analysis of different polarizations of the susceptibility tensor demonstrated that for Fe and Co the adiabatic approximation is well justified and spin dynamics in these materials has nearly pure transversal character at least up to 1 eV energy range. On the other hand, for Ni longitudinal SDF are shown to be more significant.

Using FDT, spin correlator, a major quantity characterizing SDF in metals, has been carefully evaluated by using the complete spectrum of SDF. The related effective fluctuating moment was found to be of the order of several Bohr magnetons with a significant generic contribution ($\sim 1.8\mu_B$) from excitations that involve semicore and high-energy states. A unique linear dependence of the effective fluctuating moment on the electronic population has been determined. Overall, our results indicate that the value of the effective fluctuating moment does not depend on the presence of equilibrium local moments.

Finally, we estimated the SDF correlation energy for all 3\emph{d} ferromagnets and found that it it the largest close to half-filling. It was shown that for all materials this correlation energy is comparable to the mean-field magnetic energy and, thus, it should be included in electronic structure calculations. We demonstrated that all excitations below energy of the order of 3\emph{d} electronic bandwidth are equally important and should be included on the same footing without usage of any long wavelength or atomistic approximations.

\section*{Acknowledgments}
This work was supported by the Critical Materials Institute, an Energy Innovation Hub funded by the U.S. Department of Energy (DOE). V. P. acknowledges the support from the Office of Basic Energy Science, Division of Materials Science and Engineering. V.N. and A.I. acknowledge the support from the Hamburg Centre for Ultrafast Imaging (CUI). The research was partially performed at Ames Laboratory, which is operated for the U.S. DOE by Iowa State University under contract \# DE-AC02-07CH11358.

\end{document}